\documentclass[runningheads]{llncs}

\usepackage[T1]{fontenc}
\usepackage[utf8]{inputenc}
\usepackage{graphicx}
\usepackage{rotating}
\usepackage{booktabs}

\usepackage{multirow}
\usepackage{xspace} 
\usepackage{hyperref}
\usepackage{amsmath}
\usepackage{amssymb}
\usepackage{xcolor}
\usepackage{url}
\usepackage{subcaption} 
\usepackage{microtype}  
\usepackage{tabularray}

\usepackage[inline]{enumitem} 
\newlist{inlinedesc}{description*}{1}
\setlist[inlinedesc]{%
  labelwidth=0pt,
  labelsep=0.0em,
  leftmargin=10pt,
}

\begin{document}
\newcommand{\gromacs}{\textsc{gromacs}\xspace}
\newcommand{\lammps}{\textsc{lammps}\xspace}
\newcommand{\gadget}{\textsc{OpenGadget3}\xspace}
\newcommand{\athenak}{\textsc{AthenaK}\xspace}
\newcommand{\dealiix}{\texttt{dealii-X}\xspace}
\newcommand{\tinytc}{\textsc{TinyTC}\xspace}

\newcommand{\gpumax}{Intel Data Center GPU Max\,1550\xspace}
\newcommand{\cpuspr}{Intel Xeon Platinum\,8480+\xspace}
\newcommand{\spr}{SPR\xspace}
\newcommand{\pvc}{PVC\xspace}
\newcommand{\sng}{SuperMUC-NG phase\,2\xspace}
\newcommand{\lrz}{Leibniz Supercomputing Center (LRZ)\xspace}

\newcommand{\ear}{\texttt{EAR}\xspace}
\newcommand{\pppem}{\texttt{p3em}\xspace}
\newcommand{\perf}{\texttt{perf}\xspace}
\newcommand{\xpusmi}{\texttt{xpu-smi}\xspace}
\newcommand{\rapl}{\textsc{RAPL}\xspace}

\title{Node-Level Performance and Energy Characterization of Flagship Science Applications on SuperMUC-NG Phase\, 2}
        
\subtitle{Intel Sapphire Rapids and Ponte Vecchio}
\author{
  Salvatore Cielo\inst{1}     \and
  Elmira Birang\inst{1}       \and
  Alexander P\"{o}ppl\inst{2} \and
  Sajad Azizi\inst{1}         \and
  Plamen Dobrev\inst{1}       \and
  Margarita Egelhofer\inst{1} \and
  Ivan Pribec\inst{1}         \and
  Gerald Mathias\inst{1}      
}

\authorrunning{S.\,Cielo et al.}

\institute{
  Leibniz Supercomputing Center (LRZ), Boltzmannstra{\ss}e~1,
  85748 Garching b.\,M\"{u}nchen, Germany\\
  \email{\{azizi, birang, cielo, dobrev, egelhofer, mathias, pribec\}@lrz.de}\\
  \url{https://www.lrz.de}
  \and
  Intel Deutschland GmbH, Germany\\
  \email{alexander.poeppl@intel.com}
}

\maketitle

\begin{abstract}
We present a systematic performance and energy-efficiency
characterization of five flagship scientific workloads on
\sng, the 28 PetaFLOPs system at the \lrz equipped with \cpuspr and \gpumax
(Ponte\,Vecchio, PVC) accelerators.
The selected codes span molecular dynamics
(\gromacs, \lammps), astrophysics and cosmology
(\gadget, \athenak), and finite-element PDE solvers from the
\dealiix Center of Excellence.
For each code we measure throughput and energy efficiency per compute element -- expressed as compute-elements per wall-clock second (or per Joule of consumed energy) -- on a single compute node, comparing
CPU-only (SPR) against combined CPU+GPU (SPR+PVC) configurations where available.
Energy measurements rely on lightweight code instrumentation with \pppem, or the Energy Aware Runtime (EAR) present on the system.
Our results show that
GPU offload yields $4$--$12\times$ higher throughput and up to $15\times$ better energy efficiency compared to CPU-only execution, with \lammps and \athenak benefiting most.
However, both throughput and energy gains are sensitive to problem granularity: insufficient work per GPU tile erodes the accelerator advantage, as clearly observed in \athenak at small mesh-block sizes. The power-budget utilization
is systematically lower for CPUs than it is for GPUs,
indicating that even at peak useful-work rate, most applications
running on CPUs 
leave a significant fraction of the node's thermal envelope unused.
\end{abstract}

\section{Introduction}
\label{sec:intro}

Next-generation HPC platforms combine multi-socket CPU nodes with
powerful accelerators on high-bandwidth intra-node interconnects,
creating an increasingly heterogeneous compute landscape\cite{dobrev2025benchmarkingMD}.
Understanding the performance \emph{and} energy behavior of
scientific applications on these platforms is essential for both users and system administrators;
energy in particular is both a primary driver of total cost of ownership, and a convenient measure of \emph{intensive} hardware performance for comparing different devices.

\sng at the \lrz is one of the first production systems worldwide based on Intel's general purpose data center GPU architecture, pairing \cpuspr CPUs with \gpumax.
This paper presents a systematic, multi-application characterizations of this platform, covering five codes from three scientific domains:
\begin{enumerate*}[label=(\roman*)]
\item \textbf{Molecular dynamics} {(\gromacs~\cite{Abraham2015GROMACS,GROMACS2,GROMACS3,GROMACS4},
  \lammps~\cite{Thompson2022LAMMPS})}
\item \textbf{Astrophysics and cosmology}~{
(\gadget \cite{gadget},
 \athenak \cite{Stone2026AthenaK})
        }
  \item \textbf{PDE solvers} (selected kernels from the
      \dealiix Center of Excellence~\cite{dealIIX}).
\end{enumerate*}

The codes present a variant degree of optimization and GPU utilization; we selected them based on usage cases and not efficiency. 
For each code we adopt \emph{holistic} node-level metrics for throughput and energy efficiency, enabling fair, domain-agnostic comparison between CPU-only and CPU+GPU execution modes where available. The study directly informs practitioners migrating workloads to PVC-equipped systems and provides feedback to hardware and software vendors on real application behavior.

\section{System Description}
\label{sec:system}

\sng is hosted at the \lrz (Garching, Germany). Its 240 compute nodes are each equipped with 
\begin{enumerate*}[label=(\roman*)]
    \item two \cpuspr (codename Sapphire Rapids, 56 cores, 512\,GB DDR5; SPR in the following),
    \item four \gpumax (codename Ponte\,Vecchio, 128\,GB HBM2e each, $\sim$52\,TFLOP/s FP64, henceforth PVC),
    \item Intel XeLink interconnect within the node.
\end{enumerate*} 

With a Thermal Design Power (TDP) of $350$\,Watt for SPR and a nominal $600$\,Watt for PVC (power-capped to $450$\,Watt on this system), the single-node power budget used below is $W_{\mathrm{TDP}} = 2500$\,Watt.

All the benchmarks and their dependencies are built with the latest generation of Intel oneAPI compilers on the system ($v2025.3.0$), and Intel MPI ($v2021.17.0$). In this stack, the large compute cards default to be subdivided into two logical tiles each (explicit scaling), to which Intel MPI can pin tasks sequentially. Thus the simplest usage model to target all devices is to run in parallel with $8$ MPI tasks per node. We specify below when a different strategy is used.

\section{Methodology}
\label{sec:method}

All measurements are \emph{single-node}, isolating the per-node
performance and energy cost without network effects.
Throughput and specific energy-efficiency metrics are computed as:
\begin{equation}
  \text{Throughput: } T = \frac{N_{\text{elements}} \cdot N_{\text{steps}}}{t_{\text{wall}}}
  \qquad
  \text{Energy efficiency: } E = \frac{N_{\text{elements}} \cdot N_{\text{steps}}}{e_{\text{node}}}
\end{equation}
where $N_{\text{elements}}$ is the number of compute elements
(atoms, mesh zones, degrees of freedom), $N_{\text{steps}}$ is
the number of time steps executed, $t_{\text{wall}}$ is elapsed
wall-clock time, and $e_{\text{node}}$ is total node energy
consumption.

As shown in previous work~\cite{cielo2025syclenergyefficientnumericalastrophysics}, we observe that the energy-efficiency definition is also valid for larger node counts, being the ratio of two extensive quantities (both the numerator and $e_{\text{node}}$ are proportional to the hardware size, e.g.\ number of used nodes). For throughput, an additional normalization by the number of used nodes would be necessary.

For \gromacs, energy readings are obtained via
\ear~\cite{Corbalan2023EAR}, which wraps Intel Running Average Power Limit (\rapl) -- CPU package + DRAM --
and GPU power counters, yielding integrated energy measures over the full workload.
For all other codes, consumed node energy is sampled using \pppem\footnote{\url{https://github.com/svtcli/p3em}}, which offers finer time granularity (100\,ms
resolution for all benchmarks). \pppem provides several energy meter options to adapt to different systems, vendors, and varying levels of user access to hardware counters. In this work, CPU energy is read via the Linux \perf interface --
which in turn \footnote{\url{https://www.intel.com/content/www/us/en/developer/articles/technical/software-security-guidance/advisory-guidance/running-average-power-limit-energy-reporting.html}} reads \rapl Model Specific Registers (MSRs) -- and GPU energy via \xpusmi.
A comparison between \pppem and \ear on the same data (from the \athenak benchmark) is discussed in Sect.~\ref{sec:earvsp3em}.

\section{Benchmarks and Results}
\label{sec:results}

Table~\ref{tab:codes} provides an overview of all five benchmarked codes, their offload strategies, throughput metrics, and single-node peak results.
The following subsections describe each benchmark and discuss
the corresponding measurements.

\begin{table}[ht!]
 \caption{Summary of benchmarked applications and single-node peak results. Abbreviations:
  \ear=\,Energy Aware Runtime~\cite{Corbalan2023EAR};
  \pppem =\,\url{https://github.com/svtcli/p3em}
 }
\label{tab:codes}
\newlength{\cw}
\setlength{\cw}{2.2cm}
\resizebox{\textwidth}{!}{%
\DefTblrTemplate{caption}{default}{} 
 \begin{talltblr}[
  note{a}={Only results up to $1.1\times10^7$ atoms are shown in this work; larger problems were also tested on one node.},
  note{b}={The simulation size is $2\times256^{3}$ particles, half of which are gas particles and half dark matter particles.},
  note{c}={Derived from 288 generated meshblocks, times user input of $80^3$ mesh size; the maximum refinement level is $5$.},
  note{d}={The mean simulation energy efficiency is at $\sim 660$, whereas the isolated HSML kernel reaches a peak efficiency of $\sim 6\times10^{3}$.},
 ]{
 colspec={
  Q[r,\cw]Q[c,\cw]Q[c,\cw]Q[c,\cw]Q[c,\cw]Q[c,\cw]},
  column{1} = {font=\bfseries},
  row{1} = {font=\bfseries}
}
& {\gromacs}
 & {\lammps}
 & {\gadget}
 & {\athenak}
 & {\dealiix kernels} \\
\hline
 Domain
  & Molecular dynamics
  & Molecular dynamics
  & Astrophysics and Cosmology
  & Astrophysics and Cosmology
  & PDE~/~FEM \\
 Offload strategy
  & SYCL/OpenCL native GPU~offload
  & Kokkos (SYCL)
  & Intel OpenMP offload
  & Kokkos (SYCL/OpenMP)
  & Kokkos (SYCL) \\
Throughput Metric
  & Atom-steps
  & Atom-steps
  & Particle-steps
  & Zone-cycles
  & DoF \\
Energy tool
  & \ear
  & \pppem
  & \pppem
  & \pppem
  & \pppem \\
\hline
Largest tested problem size (node)
  & {$>4.3\times10^{7}$ atoms}\TblrNote{a}
  & {$3.5\times10^{6}$ atoms}
  & $3.36\times10^{7}$ particles\TblrNote{b}
  & $\sim1.47\times10^{8}$ zones\TblrNote{c}
  & $8\times 10^{8}$ \\
Peak throughput
  & {$\sim4\times10^{8}$}
  & {${\sim}5\times10^{8}$}
  & $4.68\times10^{5}$
  & ${\sim}6\times10^{6}$
  & {${\sim}6\times10^{10}$} \\
Peak energy efficiency
  & {$\sim2\times10^{5}$}
  & {${\sim}4\times10^{5}$}
  & ${\sim}660$\TblrNote{d}
  & ${\sim}3\times10^{4}$
  & {${\sim}6\times10^{7}$} \\
\hline
\end{talltblr}}
\end{table}

\subsection{\gromacs -- Molecular Dynamics}
\label{sec:gromacs}

\gromacs~\cite{Abraham2015GROMACS,GROMACS2,GROMACS3,GROMACS4}is an open-source, highly parallel software package for running classical molecular dynamics (MD). It is capable of utilizing almost all modern hardware, including computer-center-class GPUs, offloading the major part of particle interactions onto the latter, while only small parts of the calculations required for MD, such as pressure coupling, are still carried out on the CPUs.

Here we use the
SYCL implementation of \gromacs,
with all non-bonded and bonded interactions and the atomic coordinate updates offloaded to the \pvc cards. To optimize communications among multiple GPUs, \gromacs distributes the neighboring atoms on the same processing unit (domain decomposition). The long-range electrostatics is calculated across the entire system via Particle Mesh Ewald (PME)~\cite{PME_1,PME_2}, and in this study was carried out on a separate GPU tile. Of an entire node comprised of 8 GPU tiles, 7 tiles were used for particle-particle interaction and 1 for PME calculation.

The test cases shown in the current study are explicit solvent water solutions of 15\% ethanol, with system sizes ranging from $2.5\times10^5$ to $1.1\times10^7$ atoms, increasing by roughly a factor of two at each step.

\begin{figure*}[p]
  \centering
  \begin{subfigure}[b]{0.48\textwidth}
    \centering
    \caption{\gromacs}
    \label{fig:results_gromacs}
     \includegraphics[width=\textwidth]{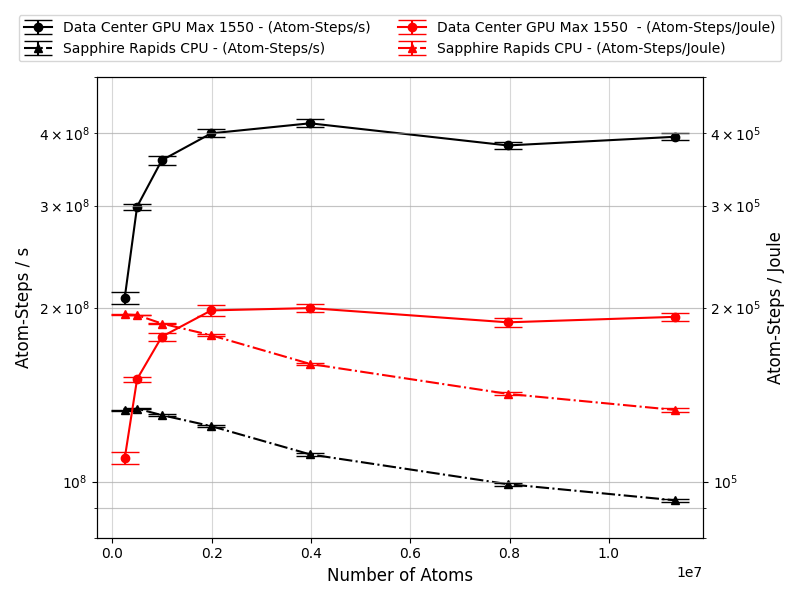}
  \end{subfigure}
  \hfill
  \begin{subfigure}[b]{0.48\textwidth}
    \centering
    \caption{\lammps (LJ fluid)}
    \label{fig:results_lammps}
 \includegraphics[width=\textwidth]{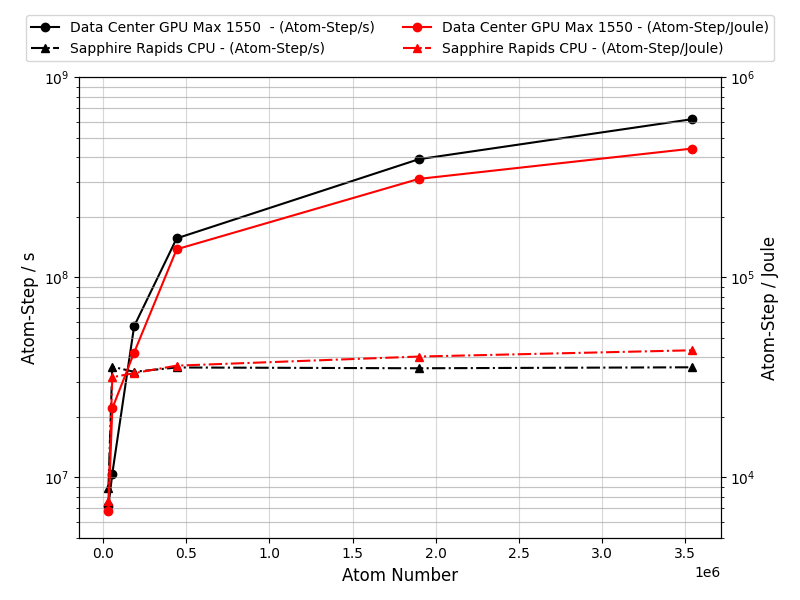}
  \end{subfigure}

  \vspace{0.5em}

  \vspace{0.25em} \hrule \vspace{0.25em}

  \begin{subfigure}[b]{0.48\textwidth}
    \centering
    \caption{\gadget}
    \label{fig:results_gadget}
\includegraphics[width=\textwidth,alt={\gadget}]{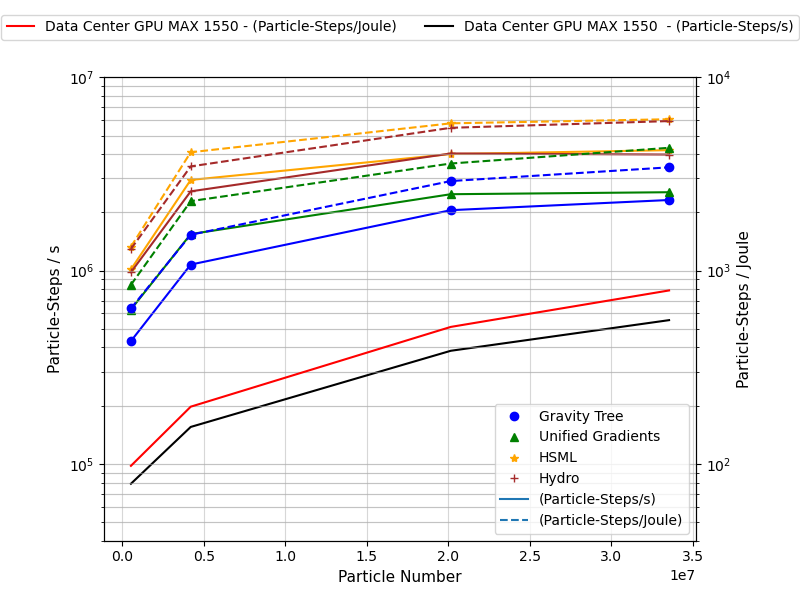}

  \end{subfigure}
  \hfill
  \begin{subfigure}[b]{0.48\textwidth}
    \centering
    \caption{\dealiix matrix-free kernels}
    \label{fig:results_dealiix}
    \includegraphics[width=\textwidth]{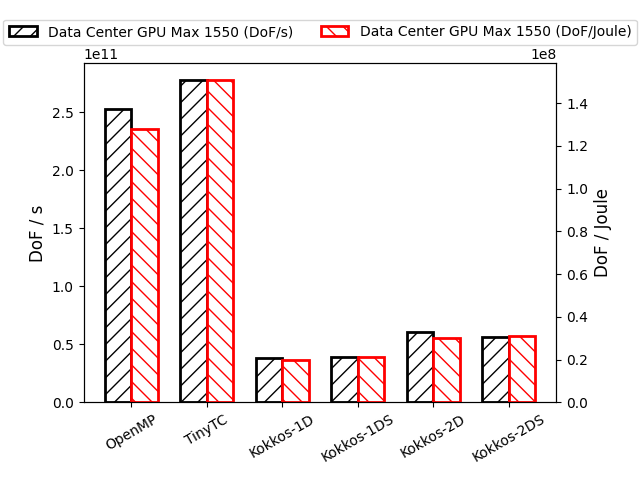}
  \end{subfigure}

  \vspace{0.25em} \hrule \vspace{0.25em}

  \begin{subfigure}[b]{0.495\textwidth}
    \centering
    \caption{\athenak (GW150914 BBH merger)}
    \label{fig:results_athenak}
     \includegraphics[width=\textwidth]{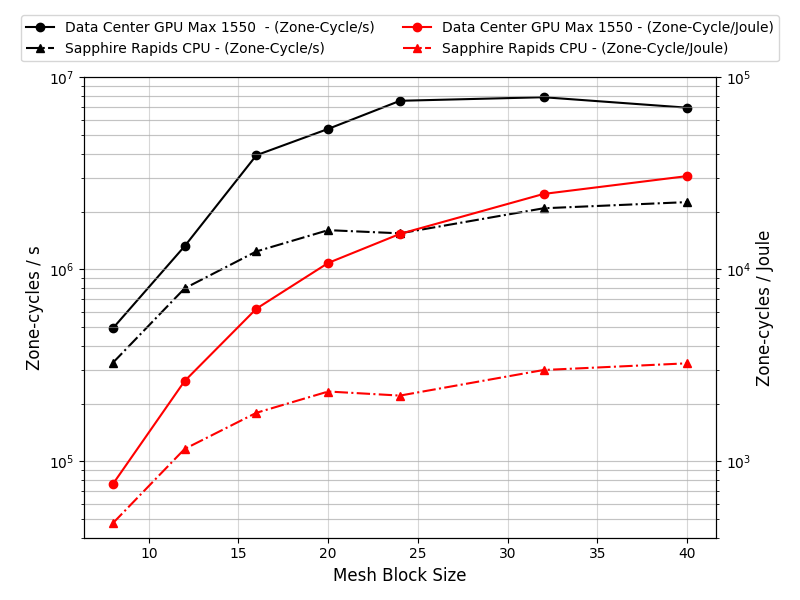}
  \end{subfigure}
  \hfill
  \begin{subfigure}[b]{0.495\textwidth}
    \centering
    \caption{\pppem-\ear comparison (\athenak data)}
    \includegraphics[width=\textwidth]{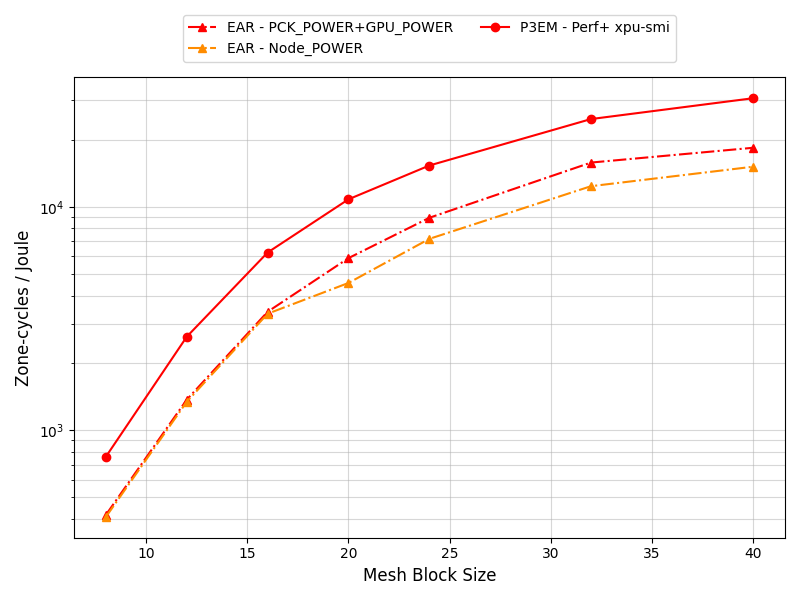}
    
    \label{fig:results_earvsp3em}
  \end{subfigure}

  \caption{Panels (a--e) show single-node performance (black, left axes) and energy efficiency (red, right axes) for the five benchmarks on
           \sng (\spr CPU-only vs.\ \spr\,+\,\pvc, where applicable).
           Solid circles / filled bars: \pvc GPU;
           dashed triangles: \spr CPU only (where applicable). 
           Panel~(\subref{fig:results_gadget}) also includes the measured values for several kernels.
           Panel~(\subref{fig:results_earvsp3em}) compares \athenak GPU
           energy efficiency across three measurement
           strategies (see Sect.~\ref{sec:earvsp3em}).}
  \label{fig:results}
\end{figure*}

Figure~\ref{fig:results}(\subref{fig:results_gromacs}) shows atom-steps per second and per Joule as a function of system size. Each configuration was run 8 times; error bars in the figures represent one standard deviation.
The GPU sustains $\sim4\times10^8$\,atom-steps/s
across all tested system sizes, approximately ${\sim}4\times$
above the \spr CPU-only trajectory.
The energy-efficiency picture is more nuanced: the GPU plateaus at
$\sim2\times10^5$\,atom-steps/J for large systems, while the
CPU efficiency \emph{decreases} with system size (dropping from
$\sim2\times10^5$ at small sizes to $\sim8\times10^4$ at
$1.1\times10^7$ atoms), likely due to cache-pressure effects.
The shape of the occupation curves is similar across throughput and energy, 
but with a clear shift: while the GPU raw throughput remains always higher, energy favors the CPU alone for smaller configurations.

\subsection{\lammps\ -- Molecular Dynamics}
\label{sec:lammps}

\lammps~\cite{Thompson2022LAMMPS} is a classical molecular dynamics code
with broad hardware support via its Kokkos package, here used with the
SYCL backend to target the \pvc cards.
The test case is a three-dimensional Lennard-Jones (LJ) fluid; system
sizes range from $3.2\times10^4$ to $3.5\times10^6$ atoms, again with the compute
device size fixed at one node.

Figure~\ref{fig:results}(\subref{fig:results_lammps}) shows that both GPU throughput and energy efficiency
increase steeply up to $\sim4\times10^5$ atoms and then plateau, reaching
$\sim5\times10^8$\,atom-steps/s and $\sim4\times10^5$\,atom-steps/J respectively.
The CPU throughput is nearly flat across all system sizes at
$\sim4\times10^7$\,atom-steps/s, while its energy efficiency remains at
$\sim4\times10^4$\,atom-steps/J, about ${\sim}10\times$ below the GPU at large sizes.
Notably, at the smallest system size the CPU and GPU energy efficiencies are
comparable, with the GPU becoming dominant only once sufficient parallelism is exposed. Here both energy and throughput convey rather similar information
(save a slight GPU decrease versus a slight CPU increase at large sizes), hinting that \lammps may have no hidden energy-consumption bottlenecks, making it one of the most efficient of the presented benchmarks.

\subsection{\gadget\ -- Large-Scale Cosmological Simulations}
\label{sec:gadget}

\gadget~\cite{gadget} is a large-scale cosmological hydrodynamics and N-body dynamics simulation code. It is highly parallel, utilizing MPI, OpenMP threads and either OpenACC or Intel OpenMP offload for computations on a GPU, making it portable to virtually all hardware systems today. In our test \gadget offloads the following kernels for computation on the GPU: gravitational tree, smoothing lengths (HSML), conduction, hydrodynamics, and density; while the rest are processed on the CPU. Communications between the CPU and GPU are kept at minimum, where only the updated values that are needed are transported. 

Making use of the hierarchical timer structure~\cite{gadgetspacetimers}, we have collected energy information from the whole simulations as well as some of \gadget's kernels -- gravitational tree, hydrodynamical acceleration, calculating smoothing length and calculating density. For our setup we run on one node, with 8 MPI tasks and 8 GPU tiles (2 tiles per GPU on the node) and 28 CPU threads per task, thus using all CPU cores and all GPUs. Figure~\ref{fig:results}(\subref{fig:results_gadget}) shows the throughput and energy efficiency for different resolution simulations. Both increase for all kernels with increasing resolution, as expected for \gadget~\cite{gadgetscaling}. In a further study we plan to increase the resolution to fill the node to maximum available memory and observe how that affects the performance.

\subsection{\athenak\ -- Binary Black-Hole Merger}
\label{sec:athenak}

\athenak~\cite{Stone2026AthenaK} is a performance-portable astrophysical
MHD code built on the Kokkos abstraction layer, here compiled with the
SYCL backend to target the \pvc cards.
The test case is the GW150914 binary black-hole merger
problem~\cite{Abbott2016GW150914}; the key tuning parameter is the
mesh-block edge length, which controls the granularity of work dispatched
to each GPU tile.

Figure~\ref{fig:results}(\subref{fig:results_athenak}) shows zone-cycles per second and per joule
as a function of mesh-block size.
The GPU configuration achieves peak throughput of
$\sim6\times10^6$\,zone-cycles/s at mesh-block edge lengths $\ge 24$,
roughly ${\sim}12\times$ higher than the \spr CPU-only peak.
From an energy-efficiency perspective the GPU reaches
$\sim3\times10^4$\,zone-cycles/J at large block sizes,
versus $\sim2\times10^3$\,zone-cycles/J for the CPU,
a ${\sim}15\times$ advantage.
Notably, at very small mesh-block edge lengths ($\le 10$) the GPU
throughput advantage narrows significantly, suggesting that
launch overhead and occupancy effects dominate at coarse granularity.

\subsection{\dealiix Kernels -- Finite-Element Solvers}
\label{sec:dealiix}

Scientific codes based on finite-element methods are a major consumer of compute cycles in the HPC environment.
Notable applications used in production on the \sng system at LRZ include SeisSol for earthquake simulations~\cite{uphoff2017extreme},
the HyTeG operator framework used in large-scale mantle convection~\cite{bohm2025code},
or the ExaDG solver for incompressible fluid mechanics and acoustics~\cite{arndt2020exadg}.
An underlying theme of these solvers is the matrix-free evaluation of the multi-dimensional
spatial PDE operators using either discontinuous or continuous Galerkin approaches.
The use of tensor-product polynomial basis functions of high order, combined with
with algebraic sum-factorization techniques provides these methods with a high arithmetic intensity well-suited for modern compute-heavy CPU and GPU architectures.

As part of an ongoing collaboration in the \dealiix Center of Excellence
we investigated the performance and energy usage of matrix-free kernels on 
the Intel \pvc GPU using different offloading frameworks including Kokkos and OpenMP \footnote{https://github.com/dealii-X/benchmarks}.
We also investigate the performance attainable using the Tiny Tensor Compiler \footnote{https://intel.github.io/tiny-tensor-compiler/}, a tensor language compiler from Intel.

The Bake-off Kernel 1 (BK1) is a synthetic test from the CEED Bake-off Problems \footnote{https://ceed.exascaleproject.org/bps/}.
It corresponds to the action of the mass matrix, $B u = f$, where $B$ is the mass matrix for a discretized scalar PDE in 3-d.
Specifically, we focus on the case with $p = 2$ (quadratic polynomials), $(p+1)^3$ nodal degrees of freedom per element and $q = p + 2$ quadrature points per spatial direction (64 quadrature points total). 
The setup consists of 100 billion elements per GPU stack, with the degrees of freedom (DoFs) stored in the E-vector format (element-local DoFs).
The evaluation of the mass matrix action involves input and output vectors and the metric tensor, in FP32 precision, giving a total memory consumption of \(10^8 \cdot (2 \cdot 3^3 + 4^3) / 1024^3 \approx 44\)~GB.
To saturate the node, 8 parallel instances of the benchmark are launched simultaneously via \texttt{mpirun}. 
This ensures a fair one-to-one pinning of processes to GPUs.
Energy consumption was captured using the p3em energy-meter API, configured to sample at 100 Hz. 
The reported performance and energy consumption were recorded as an average of 100 repeated kernel invocations, to guarantee sufficient elapsed time of the benchmark.
Originally, the benchmark programs were configured to measure the fastest single kernel invocation (i.e. minimum time), which resulted in high uncertainty when translated to the energy measurement.

The baseline version is implemented in Kokkos; four variants named 1D, 1DS, 2D, and 2DS, correspond to different thread-mapping strategies (see \cite{swirydowicz2019acceleration} for a discussion).
The OpenMP variant is derived from the serial kernel and uses the descriptive construct, \texttt{\#pragma omp target teams loop} on the outer loop across elements, giving the compiler the freedom of automatic thread mapping and parallelization. 
Finally, the TinyTC approach used manual cross-element vectorization to maximize SIMD efficiency.

Figure~\ref{fig:results}(\subref{fig:results_dealiix}) compares the performance and energy efficiency of the different implementations. 
The OpenMP and TinyTC variants exhibit higher performance and better energy usage. 
A similar pattern of the two metrics is observed across different variants, hinting that the instant power draw remained roughly constant, thus energy mainly correlates with total runtime.

\section{Discussion}
\label{sec:discussion}
\subsection{Complementarity of throughput and energy efficency}

Across the codes evaluated, the \gpumax
consistently outperforms the \cpuspr CPU in raw
throughput, with gains ranging from ${\sim}4\times$ (\gromacs)
to ${\sim}12\times$ (\athenak) depending on the application.
Energy efficiency is more application-dependent:
\athenak achieves the largest GPU energy advantage (${\sim}15\times$ at large mesh-block sizes),
followed by \lammps (${\sim}10\times$),
while \gromacs shows a more modest gain that disappears entirely
at small system sizes, where the CPU alone is equally or more
energy-efficient. 

\gadget shows the smallest energy efficiency, which could
increase with
simulation size and will be the subject of further study. Looking at the energy efficiency of isolated kernels (Fig.~\ref{fig:results}(\subref{fig:results_gadget})), we see that, for example, the HSML calculation reaches higher values of $\sim 6\times10^{3}$, about 10 times greater than the average for the whole simulation.

A key observation is the sensitivity of \emph{both} throughput
\emph{and} energy efficiency to device occupancy, as clearly visible in the \athenak results (Fig.~\ref{fig:results}(\subref{fig:results_athenak})):
insufficient work per GPU kernel invocation leads to underutilization
of PVC's massive vector width.
Practitioners porting codes to PVC-equipped systems should therefore
ensure that tile-local problem sizes are large enough to saturate
the GPU before expecting full throughput and energy benefits.

Where 
throughput and energy efficiency
track each other closely -- as in \lammps across its
scaling range, or across the \dealiix kernel variants -- instantaneous power is roughly constant and energy reduces to a runtime proxy.

\begin{figure*}[!t]
  \centering
    \label{fig:efficiency}
     \includegraphics[width=\textwidth]{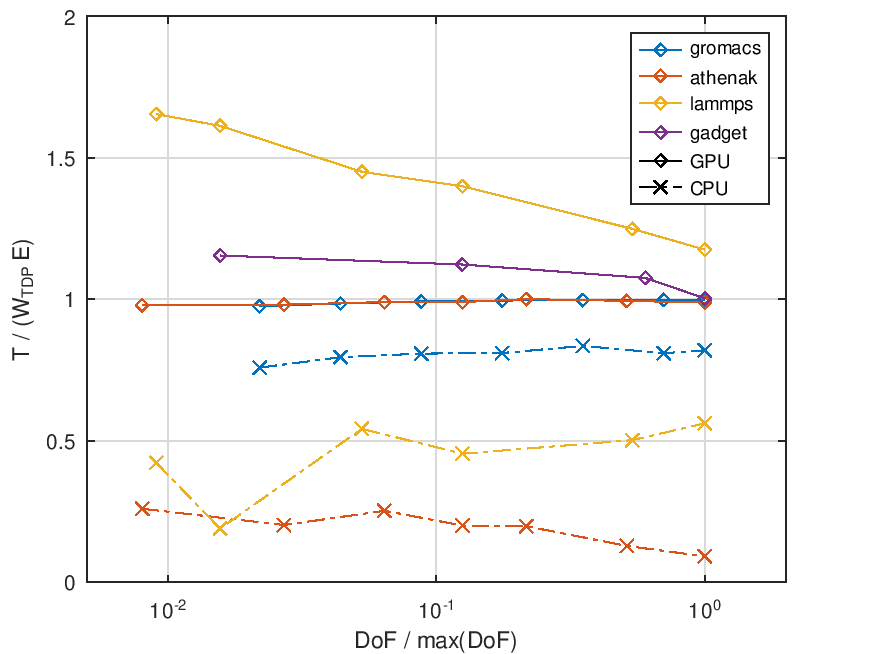}
         \caption{Power-budget parameter: $T/(W_{TDP}\,E)$ as a function of normalized degrees of freedom, for all runs except \dealiix (for which no workload size study was conducted). GPU runs are stable or converge around unity, indicating an efficient usage of power budget. CPU runs stabilize, but on values different from unity, indicating a systematic loss of power outside application computation.}
    \label{fig:budget}
\end{figure*}

In Figure \ref{fig:budget} we plot the \emph{power-budget parameter} (defined as the ratio of throughput $T$ and energy efficiency $E$, after normalizing the latter by the thermal design power $W_{TDP}$) as a function of the normalized degrees of freedom (which track workload size) for the above runs (except for \dealiix where no workload size study was conducted).

A power budget of value unity means that $E$ and $T$ convey essentially the same information, and $W_{TDP}$ is an appropriate model of power consumption. As we see, this is the case for two of the most efficient codes: \gromacs and \athenak, when running on GPU. The other codes seem to  converge or approach unit value at large DoF, suggesting that larger workload converge to optimal power budget utilization. 
At smaller value, the energy efficiency at the denominator is lower, suggesting that the GPU baseline power consumption is not fully amortized for smaller workloads.

For the CPU-only runs, despite having used the reduced $W_{TDP}=700 W$ for the two combined \spr sockets alone,
the power budget is systematically lower than unity. 
In reality, the lines do not diverge much from a stable
value; meaning that for large workload sizes (but the CPUs are more easily filled even for lower DoF) $T$ and $E$ also track each other to some extent. The deviation from unity seem to derive from a systematic fraction of the CPU power being spent on other tasks than the application computation; in other words, the $W_{TDP}$ is an unrealistically optimistic consumption model for CPUs.

As noted previously, the code ability to use the device EUs efficiently matters:
the energy consumption is dominated by compute-intense kernels, whereas wallclock is dominated by memory- or communication-bound kernels that contribute less to the energy total than their runtime share would
suggest.
As a result, two runs with similar throughput can show markedly
different aggregate energy efficiency if their internal kernel mixes
differ, and the same simulation can shift in efficiency across problem
sizes purely because the relative time fractions of its kernels change.
The \gadget breakdown
(Fig.~\ref{fig:results}(\subref{fig:results_gadget}))
illustrates this directly:
despite kernels such as HSML showing very high energy efficiency, the whole simulation has a lower overall power budget utilisation, indicating that the aggregate is dragged down by other kernels with substantially lower
per-Joule efficiency that dominate the runtime.
Together, these effects suggest that energy efficiency is a more
sensitive probe of intra-node bottlenecks than throughput alone, and
that codes with similar throughput scaling can have substantially
different optimization headroom in the energy dimension.

\subsection{\ear vs \pppem Comparison}
\label{sec:earvsp3em}

Figure~\ref{fig:results}(\subref{fig:results_earvsp3em}) shows the \athenak GPU energy efficiency measured with three configurations: 
\begin{enumerate*}[label=(\roman*)]
 \item \pppem (\perf+\,\xpusmi),
 \item \ear with CPU package\,+\,GPU power,
 \item \ear with full node power.
\end{enumerate*}

All three curves follow the same trend with mesh-block size,
confirming consistency between tools.
However, \pppem systematically reports higher efficiency than both \ear
variants, with the gap most pronounced at small block sizes.
We attribute this offset to initialization and finalization energy
captured by \ear's job-level accounting but absent from \pppem's
tighter measurement window, which brackets only the compute phase or any desired region of interest with low time latency ($\lesssim 10$ ms).
The \texttt{Node\_POWER} curve sits below \texttt{PCK\_POWER+GPU\_POWER},
as expected: integrating all node components (memory, interconnect,
auxiliary boards) increases the denominator without contributing
to useful computation.
These findings motivate the use of \pppem for the remaining benchmarks,
where phase-accurate energy attribution is essential; a detailed
account of \pppem's design and advantages is the subject of
forthcoming work (Cielo et al., in prep).

\section{Conclusions and Future Work}
\label{sec:concl}

We have presented single-node performance and energy-efficiency
measurements for five production-level scientific workloads on \sng, demonstrating both the strengths and the
sensitivity of the Intel SPR+PVC platform to workload granularity.
Our unified metric approach (compute-elements per second and per
Joule) enables cross-domain comparison and provides actionable
guidance for users migrating to PVC-equipped systems. The energy efficiency is especially significant, in that it allows for comparison between intensive performance (i.e., less dependent on the node/device \emph{size} than the raw throughput) in a data-driven fashion (as opposed to more model-dependent reports such as roofline analysis). In some cases the two metrics provide similar information, as in \lammps (Fig.~\ref{fig:results}(\subref{fig:results_lammps})) and \dealiix kernels (Fig.~\ref{fig:results}(\subref{fig:results_dealiix})); in others, energy efficiency is relatively higher on the CPU (e.g.\ \gromacs, Fig.~\ref{fig:results}(\subref{fig:results_gromacs})) or on the GPU (e.g.\ \athenak, Fig.~\ref{fig:results}(\subref{fig:results_athenak})).

Future work will extend the study to multi-node configurations to
assess the interplay of intra-node GPU performance with the system's network fabric, and will include 
results from further benchmark applications, and comparison with a broader hardware base.

\section*{Acknowledgements}

Ivan Pribec acknowledges Carsten Uphoff (Intel) for his work on the optimized TinyTC kernels and his expertise in performance optimization. Special thanks also go to collaborators Eney Soydan and Martin Kronbichler (RUB) for their partnership on the \dealiix project.

\bibliographystyle{splncs04}   
\bibliography{bib}

@InProceedings{dobrev2025benchmarkingMD,
author="Dobrev, Plamen
and Mathias, Gerald",
editor="Weiland, Mich{\`e}le
and Neuwirth, Sarah
and Kruse, Carola
and Weinzierl, Tobias",
title="Benchmarking of GPU Performance Saturation on Accelerated Cluster Nodes via Molecular Dynamics Software Packages",
booktitle="High Performance Computing. ISC High Performance 2024 International Workshops",
year="2025",
publisher="Springer Nature Switzerland",
address="Cham",
pages="115--126",
isbn="978-3-031-73716-9"
}

@misc{cielo2025syclenergyefficientnumericalastrophysics,
      title={{SYCL} for Energy-Efficient Numerical Astrophysics: the case of DPEcho}, 
      author={Salvatore Cielo and Alexander Pöppl and Ivan Pribec},
      year={2025},
      eprint={2508.14117},
      archivePrefix={arXiv},
      primaryClass={astro-ph.IM},
      url={https://arxiv.org/abs/2508.14117}, 
}

@article{PME_1,
    author = {Darden, T. and York, D. and Pedersen, L.},
    title = {Particle Mesh Ewald: An N·log(N) Method for Ewald Sums in Large Systems.},
    journal = {J. Chem. Phys.},
    volume = {98},
    pages = {10089--10092},
    year = {1995}
}

@article{PME_2,
    author = {Essmann, U. and Perera, L. and Berkowitz, M. and Darden, T. and Lee, H. and Pedersen, L.},
    title = {A Smooth Particle Mesh Ewald Method.},
    journal = {J. Chem. Phys.},
    volume = {103},
    pages = {8577--8592},
    year = {1995}
}

@article{Corbalan2023EAR,
  author = {Corbalan, Julita and others},
  title = {EAR: Energy Management Framework for Supercomputers},
  journal = {International Journal of High Performance Computing Applications},
  volume = {37},
  number = {5},
  pages = {528--548},
  year = {2023},
  doi = {10.1177/10943420231179036},
  url = {https://github.com/eas4dc/EAR}
}

@article{Abraham2015GROMACS,
  author = {Abraham, Mark J. and Murtola, Teemu and Schulz, Roland and Páll, Szilárd and Smith, Jeremy C. and Hess, Berk and Lindahl, Erik},
  title = {{gromacs}: High performance molecular simulations through multi-level parallelism from laptops to supercomputers},
  journal = {SoftwareX},
  volume = {1--2},
  pages = {19--25},
  year = {2015},
  doi = {10.1016/j.softx.2015.06.001}
}

@article{GROMACS2,
    author = {S. Páll and M. J. Abraham and C. Kutzner and B. Hess and E. Lindahl},
    title = {Tackling Exascale Software Challenges in Molecular Dynamics Simulations with
GROMACS},
    journal = {In S. Markidis and E. Laure (Eds.), Solving Software Challenges for Exascale},
    year = {2015},
    doi = {10.1007/978-3-319-15976-8_1}
}

@article{GROMACS3,
    author = {S. Pronk and S. Páll and R. Schulz and P. Larsson and P. Bjelkmar and R. Apostolov and M. R. Shirts and J. C. Smith and P. M. Kasson and D. van der Spoel and B. Hess, E. Lindahl},
    title = {GROMACS 4.5: a high-throughput and highly parallel open source molecular
simulation toolkit},
    journal = {Bioinformatics},
    year = {2013},
    doi = {10.1093/bioinformatics/btt055}
}

@article{GROMACS4,
    author = {B. Hess and C. Kutzner and D. van der Spoel and E. Lindahl},
    title = {{GROMACS 4}: Algorithms for highly efficient, load-balanced, and scalable
molecular simulation},
    journal = {J. Chem. Theory Comput.},
    year = {2008},
    doi = {10.1021/ct700301q}
}

@article{Thompson2022LAMMPS,
  author = {Thompson, Aidan P. and Aktulga, H. Metin and Berger, Richard and Bolintineanu, Dan S. and Brown, W. Michael and Crozier, Paul S. and {in 't Veld}, Pieter J. and Kohlmeyer, Axel and Moore, Stan G. and Nguyen, Trung Dac and Shan, Ray and Stevens, Mark J. and Tranchida, Julien and Trott, Christian and Plimpton, Steven J.},
  title = {LAMMPS -- A flexible simulation tool for particle-based materials modeling},
  journal = {Computer Physics Communications},
  volume = {271},
  pages = {108171},
  year = {2022},
  doi = {10.1016/j.cpc.2021.108171}
}

@article{Stone2026AthenaK,
  author = {Stone, James M. and Mullen, Patrick D. and Fielding, Drummond and Grete, Philipp and Guo, Minghao and Kempski, Philipp and Most, Elias R. and White, Christopher J. and Wong, George N.},
  title = {AthenaK: A Performance-portable Version of the Athena++ Adaptive Mesh Refinement Framework},
  journal = {The Astrophysical Journal Supplement Series},
  volume = {283},
  number = {1},
  pages = {27},
  year = {2026},
  doi = {10.3847/1538-4365/ae3717},
  archivePrefix = {arXiv},
  eprint = {2409.16053}
}

@article{Abbott2016GW150914,
  author = {Abbott, B.P. and others and {LIGO Scientific Collaboration and Virgo Collaboration}},
  title = {Observation of Gravitational Waves from a Binary Black Hole Merger},
  journal = {Physical Review Letters},
  volume = {116},
  number = {6},
  pages = {061102},
  year = {2016},
  doi = {10.1103/PhysRevLett.116.061102}
}

@article{gadget,
  author = {Shukla, Nitin and others},
  title = {EuroHPC SPACE CoE: Redesigning Scalable Parallel Astrophysical Codes for Exascale },
  journal = {CF '25 Companion: Proceedings of the 22nd ACM International Conference on Computing Frontiers: Workshops and Special Sessions},
  pages = {177-184},
  year = {2025},
  doi = {10.1145/3706594.3728892}
}

@misc{gadgetscaling,
  author = {Hammer, N. and others},
  title = {Extreme Scale-out SuperMUC Phase 2 - lessons learned },
    year={2016},
      eprint={1609.01507},
      archivePrefix={arXiv},
      url={https://arxiv.org/abs/1609.01507}
}

@misc{gadgetspacetimers,
      title={SPACE-Timers -- A Stack-Based Hierarchical Timing System for C++ }, 
      author={ Karademir, Geray S. ; Dolag, Klaus },
      year={2026},
      eprint={2603.01618},
      archivePrefix={arXiv},
      url={https://arxiv.org/abs/2603.01618}, 
}

@misc{dealIIX,
  author = {{deal.II-X Centre of Excellence}},
  title = {deal.II-X Centre of Excellence},
  url = {https://www.dealii-x.eu/},
  note = {Formal reference if available (deliverable / paper)}
}

@misc{tinytc,
  author = {{Intel Corporation}},
  title = {tinytc -- Tiny Tensor Compiler},
  url = {https://github.com/intel/tiny-tensor-compiler}
}

@incollection{arndt2020exadg,
  title={ExaDG: High-order discontinuous Galerkin for the exa-scale},
  author={Arndt, Daniel and Fehn, Niklas and Kanschat, Guido and Kormann, Katharina and Kronbichler, Martin and Munch, Peter and Wall, Wolfgang A and Witte, Julius},
  booktitle={Software for exascale computing-SPPEXA 2016-2019},
  pages={189--224},
  year={2020},
  publisher={Springer}
}

@article{bohm2025code,
  title={Code generation and performance engineering for matrix-free finite element methods on hybrid tetrahedral grids},
  author={B{\"o}hm, Fabian and Bauer, Daniel and Kohl, Nils and Alappat, Christie L and Th{\"o}nnes, Dominik and Mohr, Marcus and K{\"o}stler, Harald and R{\"u}de, Ulrich},
  journal={SIAM Journal on Scientific Computing},
  volume={47},
  number={1},
  pages={B131--B159},
  year={2025},
  publisher={SIAM}
}

@inproceedings{uphoff2017extreme,
  title={Extreme scale multi-physics simulations of the tsunamigenic 2004 sumatra megathrust earthquake},
  author={Uphoff, Carsten and Rettenberger, Sebastian and Bader, Michael and Madden, Elizabeth H and Ulrich, Thomas and Wollherr, Stephanie and Gabriel, Alice-Agnes},
  booktitle={Proceedings of the international conference for high performance computing, networking, storage and analysis},
  pages={1--16},
  year={2017}
}

@article{swirydowicz2019acceleration,
  title={Acceleration of tensor-product operations for high-order finite element methods},
  author={{\'S}wirydowicz, Kasia and Chalmers, Noel and Karakus, Ali and Warburton, Tim},
  journal={The International Journal of High Performance Computing Applications},
  volume={33},
  number={4},
  pages={735--757},
  year={2019},
  publisher={SAGE Publications Sage UK: London, England}
}

\end{document}